\documentclass[aps,pre,reprint, superscriptaddress]{revtex4-2}
\usepackage{blindtext}
\usepackage[colorlinks=true,allcolors=blue]{hyperref}

\usepackage{blindtext}
\usepackage{bm}
\def \beq{\begin{eqnarray}}
\def \eeq{\end{eqnarray}}

\newcommand{\mbf}[1]{\bm{#1}}
\usepackage{amsmath}
\usepackage{amssymb}
\usepackage{graphicx}
\usepackage{graphicx,xcolor}

\begin{document}
\title{Hydrodynamic stresses in a multi-species suspension of active Janus colloids}
\author{Gennaro Tucci}
\email{gennaro.tucci@ds.mpg.de}
\altaffiliation{These authors contributed equally to this work}
\affiliation{Max Planck Institute for Dynamics and Self-Organization (MPIDS), D-37077 Göttingen, Germany}

\author{Giulia Pisegna}
\email{giulia.pisegna@ds.mpg.de}
\altaffiliation{These authors contributed equally to this work}
\affiliation{Max Planck Institute for Dynamics and Self-Organization (MPIDS), D-37077 Göttingen, Germany}

\author{Ramin Golestanian}
\email{ramin.golestanian@ds.mpg.de}
\affiliation{Max Planck Institute for Dynamics and Self-Organization (MPIDS), D-37077 Göttingen, Germany}
\affiliation{Rudolf Peierls Centre for Theoretical Physics, University of Oxford, Oxford OX1 3PU, United Kingdom}

\author{Suropriya Saha}
\email{suropriya.saha@ds.mpg.de}
\affiliation{Max Planck Institute for Dynamics and Self-Organization (MPIDS), D-37077 Göttingen, Germany}

\begin{abstract}
A realistic description of active particles should include interactions with the medium, commonly a momentum-conserving simple fluid, in which they are suspended. In this work, we consider a multi-species suspension of self-diffusiophoretic Janus colloids interacting via chemical and hydrodynamic fields. Through  a systematic coarse-graining of the microscopic dynamics, we calculate the multi-component contribution to the hydrodynamic stress tensor of the incompressible Stokesian fluid in which the particles are immersed. For a single species, we find that the strength of the stress produced by the gradients of the number density field is determined by the particles' self-propulsion and chemotactic alignment, and can be tuned to be either contractile or extensile. For a multi-species system, we unveil how different forms of activity modify the stress tensor, and how non-reciprocity in hydrodynamic interactions emerges in an active binary mixture.
\end{abstract}

\maketitle

%introduction and motivations
\section{Introduction}\label{sec:Intro}
The field of soft active matter encompasses a wide range of systems from the living to the synthetic which exhibit varied collective behaviour~\cite{gompper20202020, marchetti2013hydrodynamics, ramaswamy2010mechanics,golestanian1909phoretic}. Minimal models play a crucial role in understanding how collective behaviour emerges at the macroscopic scale starting from simple interaction rules between individual particles \cite{vicsek1995novel,chate2020dry}. However, to bridge the gap between theoretical predictions and experimental observations, it is essential to incorporate the intrinsic complexity of living and active matter, and use the outcome in systematic coarse-graining and the construction of effective theories at large scales. %; with progress often occurring in incremental steps. 

A simplifying assumption in many minimal models is to ignore the effect of the medium, often a simple fluid, in which the active agents are suspended. For example, the simplest model for flocking is a collection of self-propelled active agents moving on a momentum-damping substrate, i.e. the momentum is a fast variable~\cite{chate2020dry,toner1998flocks}. When momentum is conserved, it introduces long-range hydrodynamic interactions that significantly alter the large-scale dynamics of particles suspended in a viscous fluid \cite{hatwalne2004rheology, saintillan2015theory}, producing striking effects such as active turbulence in the regime of low Reynolds number~\cite{aditi2002hydrodynamic,dombrowski2004self,wolgemuth2008collective,uchida2010synchronization,dunkel2013fluid,maitra2018nonequilibrium}. 

\begin{figure*}[t]
    \centering
    \includegraphics[width = 1.0\linewidth]{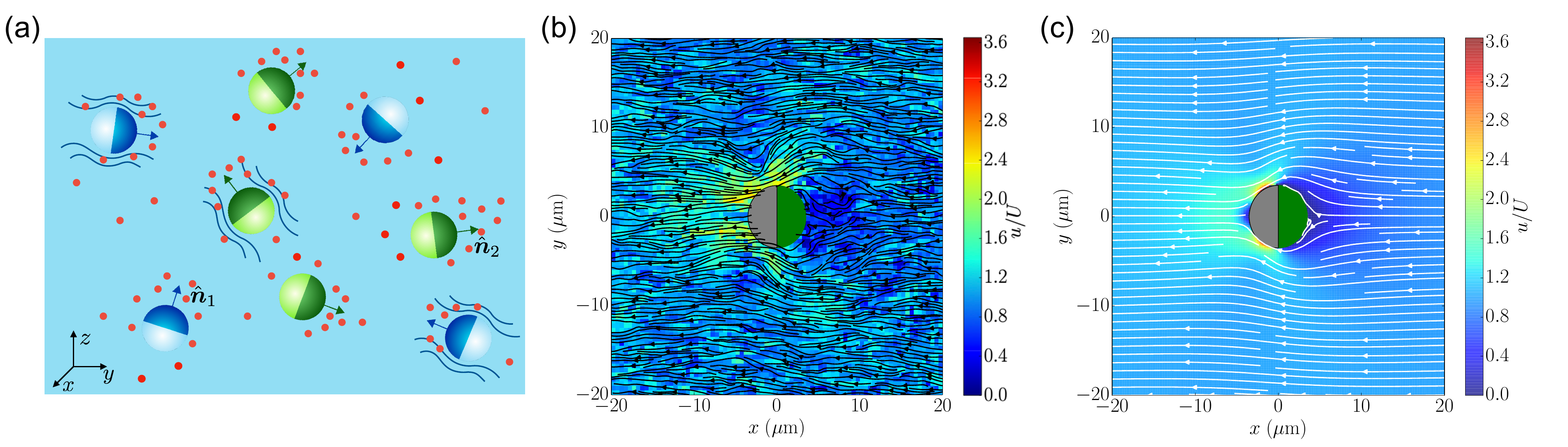}
    \caption{Panel (a): schematic of a suspension with two species of Janus colloids. Particles are immersed in a momentum-conserving fluid, they self-propel and reorient to solute concentration gradients (red dots). Phoretic and hydrodynamic interactions rule the dynamics of the colloids. Reproduced from Ref. \cite{campbell2019experimental}, we show the hydrodynamic streamlines around a catalytically active Janus particle as probed experimentally (b) and theoretically (c).}
    \label{fig:schematic}
\end{figure*}

Coupling to a fluid has so far been largely overlooked in systems with non-reciprocal interactions~\cite{soto2014,ivlev2015statistical,agudo2019active, saha2020scalar,you2020nonreciprocity,fruchart2021non}, with a few exceptions \cite{uchida2010synchronization,Nasouri2020,Nasouri_Golestanian_2020}. In these systems, which exhibit an effective breaking of action-reaction symmetry, the medium plays the important role of a momentum sink ensuring that the microscopic interactions faithfully preserve the symmetry. In this work, we fill this gap by employing the coarse-graining methodology to connect the microscopic time evolution to the large-scale dynamics, in order to systematically explore the collective behavior of a suspension of multi-species diffusiophoretic active particles with hydrodynamic interactions (see Fig.~\ref{fig:schematic}). 

Phoretic particles are synthetic colloids that respond to chemical gradients in their environment \cite{golestanian1909phoretic,illien2017fuelled}, mimicking chemotaxis \cite{berg1977physics,celani2010bacterial}. These are suspended in a fluid and catalyze the dissociation of a chemical substrate on their surface, generating solute gradients, driving diffusiophoresis, and inducing spontaneous particle motion \cite{golestanian1909phoretic}. Catalytic Janus colloids~\cite{golestanian2005propulsion,golestanian2007designing,howse2007self}, i.e. phoretic active particles with a polarity, have the ability to self-propel and align with the local chemical gradient (or in the opposite direction) \cite{saha2014clusters, saha2019pairing}. Several aspects of their ability to follow gradients have been observed experimentally, primarily at the single-particle level \cite{howse2007self,Baraban2013}. Over the past decade, they have become widely used examples of synthetic active systems, enabling both basic research \cite{ebbens2010self,das2015boundaries,simmchen2016topographical,sharan2023pair,campbell2019experimental} as well as bio-medical and environmental applications \cite{Joseph2017, soto2020onion, mundaca2020zinc, tang2020enzyme, venugopalan2020fantastic, soto2021smart, ressnerova2021efficient, mayorga2021swarming, oral2021self, ussia2021active, hortelao2020lipobots, wang2019lipase, rahiminezhad2020janus}. 
Recent studies have focused on catalytic colloid mixtures revealing non-reciprocal interactions in the dynamics of the coarse-grained densities and orientation fields \cite{tucci2024nonreciprocal}, paving the way for experimental realizations of active field theories with non-reciprocity \cite{saha2020scalar, you2020nonreciprocity}.

In this paper, we consider the dynamics of active colloids as coupled with an incompressible Stokesian fluid (low Reynolds number) \cite{happel2012low,Brady_stokes}. With the aim of obtaining an expression for the stresses in the fluid, we first characterize the contribution from a single active particle, and then extend the calculation to a collective multi-species system. We are able to calculate the active stress tensor generated by the action of the particles on the fluid and provide a microscopic interpretation of its non-equilibrium nature. Moreover, we connect the linear instabilities of the homogeneous suspensions to the hydrodynamic behavior of the swimming particles. %The rest of the paper is organized as follows. In Sec. \ref{sec:th}, the theoretical framework is laid out, and followed by the development of the fluid dynamics of the stress tensor in Sec. \ref{sec:fluidDyn}. Section \ref{sec:SingleSwimmer} describes calculations for a single Janus micro-swimmer, which is followed by the extension to collective effects for many particles presented in Sec. \ref{sec:collective}, and concluding remarks in Sec. \ref{sec:concl}. Finally, some details of the calculations are relegated to Appendices.

\section{Theoretical framework}\label{sec:th}

We consider the dynamics of a total of $N$ phoretic particles belonging to $N_s $ distinct species (Fig.~\ref{fig:schematic}a) moving in a three-dimensional fluid. There are $N_a$ particles of species $a$, each being labelled by the index $\gamma=1,\cdots,N_a$ identifying individuals of type $a$, such that $N=\sum_{a = 1}^{N_s} N_a$. The position $\bm{r}_{a}^\gamma$ and the orientation given by the unit vector $\hat{\bm{n}}_{a}^\gamma$ of the Janus colloids evolve according to the following equations:
\begin{equation}\label{eq:langevin}
\begin{aligned}
\frac{\mathrm{d}\bm{r}_{a}^\gamma}{\mathrm{d}t}&=V_a\hat{\bm{n}}_{a}^\gamma-\mu_{a} \bm\nabla c+\nu_{a}\left(\hat{\bm{n}}_{a}^\gamma \hat{\bm{n}}_{a}^\gamma-\frac{1}{3}\mathbb{I}\right)\cdot \bm\nabla c\\
&+\bm{v}(\bm r_{a}^\gamma)+\sqrt{2D_a}\,\bm{\xi}_{a}^\gamma,\\
\frac{\mathrm{d}\hat{\bm{n}}_{a}^\gamma}{\mathrm{d}t}
&=\left(\mathbb{I}-\hat{\bm{n}}_{a}^\gamma\hat{\bm{n}}_{a}^\gamma\right)\cdot(\Omega_{a}\bm\nabla c+\mathbb{W}\cdot\hat{\bm{n}}_{a}^\gamma+\sqrt{2D_{r,a}}\,\bm{\zeta}_{a}^\gamma),
\end{aligned}
\end{equation}
The fluctuating nature of Eq. \eqref{eq:langevin} is encoded in the Gaussian white noises $\bm{\xi}_{a}^\gamma(t)$ and $\bm{\zeta}_{a}^\gamma(t)$, while $D_a$ and $D_{r,a}$ are the noise amplitudes for translation and rotation, respectively. The particle self-propels with constant velocity $V_a$ along $\hat{\bm{n}}_{a}^\gamma$ as a result of an axis-symmetric catalytic coating that makes its surface chemically active and produces a local chemical gradient~\cite{golestanian1909phoretic}. It experiences a chemotactic drift as a response to external gradients of the concentration $c$ of solute particles with scalar mobility $\mu_a$ and the polarity dependent mobility $\nu_a$. It is advected by the velocity field $\bm{v}$ of the fluid at the position of the particle. The orientation experiences an angular velocity due to the fluid vorticity $\mathbb{W}=[\bm\nabla\bm{v}-( \bm\nabla\bm{v})^T]/2$ and another that tries to align it with the local chemical field via the term $\Omega_a$. $\mathbb{I}$ is the $3$-dimensional identity matrix. Note that we consider only spherical particles, and we thus neglect terms given by shape factors \cite{jeffery1922motion}. 

The chemotactic response to the gradient of the  chemical substrate $\bm \nabla c (\bm r,t)$, with which particles interact at their surface, plays a crucial role in the dynamics of the colloids and is different for each species. The sign of the chemotactic mobility $\mu_{a}$ prescribes whether the particle moves following increasing ($\mu_{a}<0$) or decreasing ($\mu_{a}>0$) gradients of the chemical density field $c$. Similarly, the coupling $\Omega_{a}$ accounts for the alignment ($\Omega_{a}>0$) or anti-alignment ($\Omega_{a}<0$) of the particle velocity along the chemical gradients \cite{tucci2024nonreciprocal}. 

The set of dynamical equations has to be combined with the evolution of the chemical field $c(\bm r, t)$, which we consider to be quasi-stationary due to a convenient separation of time scales. The equation reads
\beq
\label{eq:c}
-D_c(\nabla^2 -\kappa^2)c(\bm r,t) =\sum_{a}^{N_s} \sum_{\gamma}^{N_a} \alpha_a( \bm r)  \delta( \bm r-\bm r_{a}^\gamma(t)),
\eeq
%%%% should we specify the dimensions of the delta??
where $D_c$ is the diffusivity of the chemical, $D_c\kappa^2$ represents its spontaneous degradation rate, and $\alpha_a$ is the rate at which the $a$-th species produces ($\alpha_a>0$) or consumes ($\alpha_a<0$) the chemical \cite{golestanian1909phoretic, agudo2019active, ouazan2021non}. 

Both the chemical and the particles are immersed in a momentum-conserving, incompressible fluid of velocity $\bm v$
that is governed by the Stokes equation in the low Reynolds number regime~\cite{currie2002fundamental}
\begin{equation}
\begin{aligned}
\label{eq:stokes}
-\eta \nabla^2 v_i + \partial_i P &=  2 \pi \eta\partial_j \Sigma_{ij}, \\
\partial_i v_i &=0,
 \end{aligned}
\end{equation}
where $i,j$ denote the Cartesian components, $P$ is the pressure, $\eta$ is the viscosity of the fluid, and $ \Sigma_{ij}$ is the stress tensor produced by the particles on the fluid. We point out that the fluid not only transports the colloids but also advects the chemical. However, the latter effect produces higher order terms in our calculation, and is therefore not considered here.

The central result of our analysis is the identification of the stress tensor $ \Sigma_{ij}$ at the collective level. %\giulia{To compute it we use the profile $c$ given by Eq. \eqref{eq:c} as a boundary condition for Eq. \eqref{eq:stokes} for a single swimmer allowing us to evaluate the fluid flow generated by a single species of active particles.} 
To calculate $\Sigma_{ij}$, we begin by evaluating the contribution from a single particle. First, we consider the \textit{self-phoretic} effect of the particle's own activity on the surrounding fluid. Specifically, we derive the chemical concentration field generated by an axisymmetrically coated spherical Janus colloid. The resulting inhomogeneous distribution of the chemical field induces a slip velocity across the colloid's surface.
We use this slip velocity as a boundary condition to determine the corresponding fluid flow from Eq. \eqref{eq:stokes}. 
%In the far-field approximation, we show that the same velocity field can be recovered by replacing the slip velocity boundary condition with a point-like force dipole (stresslet), which allows us to reconstruct the stress exerted on the fluid.
Next, we evaluate the stress generated by an external chemical gradient, which arises from the collective activity of all other colloids. Finally, combining all these contributions and leveraging the linearity of Stokes flow, we extend our analysis to the collective regime.

In the following sections, we provide a detailed account of these calculations. However, we first list the key results concerning the large-scale dynamics of the system.
As we are interested in the slowest modes of the theory, we can coarse-grain the microscopic dynamics of Eqs. \eqref{eq:langevin} and reduce the description to a set of hydrodynamic equations for $\rho_a(\bm r,t)=\langle \sum_\gamma \delta(\bm r- \bm r_a^\gamma)\rangle$, i.e., the density of particles of species $a$ and the fluid velocity $\bm{v}(\bm r,t)$. The polarization field is a fast mode of the theory \cite{tucci2024nonreciprocal}, and can therefore be enslaved to the density field \cite{golestanian2012}. This allows us to close the hierarchy of moments and arrive at the continuity equation satisfied by the conserved densities $\rho_a$, namely
\beq \label{eq:rhoa}
\partial_t \rho_a + (\bm{v} \cdot \bm\nabla) \rho_a= D^{\rm eff}_a \nabla^2 \rho_a + \sum_{b}^{N_s} \chi_{ab} \bm\nabla \cdot ( \rho_a \bm\nabla \rho_b).
\eeq 
These conserved and advected density field for each species interacts with different species through the coupling constant $\chi_{ab}$, which embed non-reciprocal interactions when $\chi_{ab} \neq \chi_{ba}$ \cite{tucci2024nonreciprocal, agudo2019active}. We derive an expression for the phoretic stress tensor $\Sigma_{ij}$ of Eq. \eqref{eq:stokes}, which reads
\begin{equation}\label{eq:stresstensor}
\begin{aligned}
    \Sigma_{ij}&=\sum_{ab}\left[2\lambda_{ab}\rho_a\left(\partial_i\partial_j\rho_b-\frac{\delta_{ij}}{3}\nabla^2\rho_b\right) \right.\\ 
&\left. +\sigma_{ab}\left(\partial_i\rho_a\,\partial_j\rho_b+\partial_j\rho_a\,\partial_i\rho_b-\delta_{ij}\frac{2}{3}\bm\nabla\rho_a\cdot\bm\nabla\rho_b\right)\right].
\end{aligned}
\end{equation}
%with $i,j$ indicating the spatial components. 
to the leading order. This tensor is traceless and symmetric in spatial derivatives, while the symmetry does not extend to the species' labels. Indeed, all the coefficients of the dynamics, such as $D_a^{\text{eff}}, \chi_{ab}, \lambda_{ab}$ and $\sigma_{ab}$ depend on microscopic parameters and, $\chi_{ab}$ and $ \lambda_{ab}$ are in general not symmetric under exchange $a \leftrightarrow b$ (see Eqs. \eqref{eq:parameters_rho}, \eqref{eq:parameters_stress1}, \eqref{eq:parameters_stress2} below). We note that the two contributions in Eq.~\eqref{eq:stresstensor} are distinct only when we consider a mixture, i.e., more than a single species.  In Appendix \ref{sec:CG}, we report the detailed coarse-graining procedure for Eq. \eqref{eq:rhoa} and \eqref{eq:stresstensor}, where we omit terms that can be reabsorbed in the pressure and vanish under projections orthogonal to the gradient operator. The expression of the coefficients in terms of microscopic parameters is listed in Sec. \ref{sec:collective}, from which we primarily find that both $\lambda_{ab}$ and $\sigma_{ab}$ are proportional to the self-propulsion speed $V_a$, the chemotactic alignment coupling $\Omega_a$, and the chemical activity $\alpha_b$. We therefore derive a purely non-equilibrium stress tensor produced by the activity of the colloids, corresponding to a clear mapping between microscopic dynamics and mesoscopic description \cite{PhysRevE.89.062316,PhysRevResearch.3.013100,pisegna2025}. With our results, the mobility and activity coating of the particles can be tuned to find relevant hydrodynamic effects at large scales. For instance, in a single-species case our calculation provides a microscopic interpretation of the active stress tensor reported in models such as active Model H \cite{tiribocchi2015active}. On the other hand in the two species case, it explains how non-reciprocal interactions can be mediated by fluid flows \cite{pisegna2025}. 

In the rest of the paper, we report the detailed calculations leading to Eq. \eqref{eq:rhoa} and Eq. \eqref{eq:stresstensor}. We start with a discussion of basic concepts of fluid mechanics that are useful in the derivation of the final equation in Sec.\ref{sec:fluidDyn}. In Sec. \ref{sec:SingleSwimmer}, we present a solution for the Stokes flow of a single particle in the case where the solute evolves with a screened reaction-diffusion process as in Eq. \eqref{eq:c}. Next, we derive the solution for Janus particles reacting to external chemical gradients. In Sec. \ref{sec:collective}, we combine these two contributions to include the collective effects in the system and perform a systematic coarse-graining and moment expansion to obtain equations of motion for the number density fields. Finally, we discuss the structure of the obtained stress-tensor for single-species and multi-species active suspensions. Some details of the calculations are relegated to Appendices.

\section{Fluid dynamics: stresslet}\label{sec:fluidDyn}

A spherical swimmer of radius $R$, located at the origin of our reference frame, produces a velocity field $\bm{v}$ that satisfies the incompressible Stokes equation with viscosity $\eta$ and pressure $P$, which read
\begin{eqnarray}\label{eq:Stokes}
    &&-\eta\nabla^2{v}_i =-\partial_i P,\, \  \,\partial_i v_i = 0.
    \end{eqnarray}
The swimmer has a slip velocity $\bm{v}_s$ prescribed on its surface that leads to a self-propulsion velocity  $V_i = - \langle v_{s,i} \rangle$ \cite{happel2012low}, where the angular brackets denote averaging over the swimmer surface. The velocity field satisfies the boundary conditions     
    \begin{eqnarray}\label{eq:BC}
       &&     \lim_{r\to\infty}{v}_i = 0,\;\;\;\;
        {v}_i|_{r=R}={V}_i+{v}_{s,i},
\end{eqnarray}
where $r$ is measured from the center of the swimmer. To derive the hydrodynamic stress in the Stokes equation, we rewrite Eq.~\eqref{eq:Stokes} by introducing force singularities on the RHS of the equation \cite{happel2012low}. The effective dynamics only holds at length scales large compared to $R$.

%The building block of the derivation of the stress tensor in Eq. \eqref{eq:stresstensor} is the solution of the Stokes Eq. \eqref{eq:stokes} in terms of point-like singularities \cite{happel2012low}.
The flow around a finite-size active particle can be expressed as a multipole expansion of the surface stress \cite{PhysRevLett.112.118102}. Keeping only the lowest order in gradients, the fluid velocity becomes
\begin{equation}\label{eq:vstresslet}
    v_i=S_{kj} \partial_j G_{ki}+O\left((R/r)^3\right).
\end{equation}
The traceless symmetric tensor $S_{ij}$ represents the strength of a stresslet singularity of the Stokes equation according to
\begin{equation}\label{eq:stokesstresslet}
    -\eta\nabla^2 v_i+\partial_i P=S_{ij}\partial_j \delta(\bm{r}),
\end{equation}
while $G_{ij} = (\delta_{ij}/r+r_i r_j/r^3)/(8\eta \pi)$ is the Oseen tensor, which is the fundamental solution (Stokeslet) of the Stokes equation in the presence of a point-like force \cite{happel2012low}. We note that the stresslet is the dominant term because the contribution from a Stokeslet (decaying as $1/r$) vanishes for a swimmer that by definition does not apply a net force on the fluid. In writing Eq. \eqref{eq:stokesstresslet}, we also neglect the rotlet involving  a single spatial gradient but producing flows that decay as $R^3/r^3$ and higher order singularities involving two or more gradients. For a spherical active particle of radius $R$ with surface velocity $\bm{v}_s$, the symmetric matrix $S_{ij}$ is given by \cite{PhysRevLett.112.118102}
\begin{equation}\label{eq:Sij}
    S_{ij}=-10\pi\eta R^2\left(\langle v_{s,i}\hat{r}_j\rangle+\langle v_{s,j}\hat{r}_i\rangle\right),
\end{equation}
where $\hat{\bm{r}}$ is the normal unit vector to the surface of the particle. For the interacting swimmers that we are considering here, the slip velocity has two contributions, one produced by its own chemical activity and the second in response to chemical gradients produced by other particles. 

\section{Contribution from a single swimmer}\label{sec:SingleSwimmer}

To derive the hydrodynamic stresses produced by a collection of phoretic Janus colloids, we first consider the flow generated by a single Janus particle. The production of the chemical at the surface of each colloid generates a slip velocity in its vicinity, which modifies the fluid profile \cite{golestanian1909phoretic}. Then, we discuss the fluid flow solution of a Janus particle responding to an external chemical gradient, e.g., the one originated by all the other colloids. 

\subsection{Flow due to the chemical activity of a single Janus particle}

To compute the \emph{self} contribution, we first consider an isolated Janus colloid of radius $R$, with surface activity $\alpha$
that is axis-symmetric with respect to the unit vector $\hat{\bm n}$. If we set $\hat{\bm n}$ to coincide with the $\hat{\bm e}_{z}$ axis of our reference frame and its origin with the center of the Janus colloid, the activity will depend only on $\cos\theta$, where $\theta$ is the angle defined with respect to $\hat{\bm n}$. Assuming fast relaxation to the steady state, the substrate density $c(r,\theta)$ depends only on the distance $r>R$ from the colloid surface and the angle $\theta$. It satisfies the set of equations \cite{golestanian1909phoretic, michelin2017geometric}
\begin{equation}\label{eq:Ceq}
    \begin{aligned}
    &-(\nabla^2-\kappa^2)c(r,\theta)=0,\\
    &-D_c(\hat{\bm e}_{r}\cdot \bm\nabla c)|_{r=R}=\alpha(\theta),\\
    &\lim_{r\to\infty}c(r,\theta)=0,
    \end{aligned}
\end{equation}
where $\hat{\bm e}_{r}$ denotes the radial unit vector. Given the symmetry of the system, the solution to Eq. \eqref{eq:Ceq} can be cast as
\begin{equation}\label{eq:cone}
    c(r,\theta)=\left(\frac{R}{r}\right)^{1/2}\sum_{m=0}^\infty c_m \,K_{m+1/2}(\kappa r) \,P_m(\cos\theta),
\end{equation}
where $P_m$ denotes the Legendre polynomial of order $m$, and $K_m(x)$ represents the modified Bessel function of the second kind of degree $m$~\cite{Arfken:math}. The coefficients $c_m$ are fixed by imposing that the flux of the chemical $c$ on the surface of the Janus colloid is regulated by the activity $\alpha$ (see Eq. \eqref{eq:Ceq}):
\begin{equation}
    c_m =%-\frac{\alpha_m}{D_c\, f'_m(R)}=
    \frac{R\,\alpha_m}{D_c[(\kappa R) \,K_{m+3/2}(\kappa R)-m\,K_{m+1/2}(\kappa R)]},
\end{equation}
where $\alpha_m$ is the $m$-th coefficient in the expansion of the activity $\alpha$ in the Legendre polynomials, i.e.
\begin{equation}\label{eq:legexp}
\begin{aligned}
\alpha(\theta) &= \sum_{m=0}^\infty \alpha_m\,P_m(\cos\theta),\\
\alpha_m &= \left(m+\frac{1}{2}\right)\int_{0}^{\pi}\mathrm{d}\theta\,\sin\theta \,P_m(\cos\theta)\,\alpha(\cos\theta).
\end{aligned}
\end{equation}

The inhomogeneous chemical concentration $c$ on the surface of the colloid generates an effective slip velocity given by~\cite{anderson1989colloid} 
\begin{equation}\label{eq:vs}
\begin{aligned}
    \bm{v}_s(\theta)=&\mu(\theta)(\mathbb{I}-\hat{\bm e}_{r}\hat{\bm e}_{r})\cdot \bm\nabla c(r,\theta)|_{r=R}\\
    %=&\mu(\theta)(\mathbb{I}-\hat{\bm e}_{r}\hat{\bm e}_{r})\cdot\left[\frac{\partial c}{\partial r}\hat{\bm e}_{r}+\frac{1}{r\sin\theta}\frac{\partial c}{\partial \theta}\hat{\bm e}_{\theta}\right]_{r=R}\\
    =&-\frac{\mu(\theta)\sin\theta}{R}\,\sum_{m=0}^\infty c_m\,K_{m+1/2}(\kappa R)\, P'_m(\cos\theta)\,\hat{\bm e}_{\theta},
\end{aligned}
\end{equation}
where $\hat{\bm e}_{\theta}$ denotes the tangential unit vector on the surface of the sphere, and $\mu(\theta)$ denotes the axisymmetric mobility of the Janus colloid. The self-propulsion velocity $\bm V$ of the Janus particle can be calculated by averaging the slip velocity field over its surface, leading to \cite{anderson1989colloid,golestanian2007designing}
\begin{equation}\label{eq:Vsp}
\begin{aligned}
    \bm{V}&=-\langle\bm{v}_s\rangle \\
    %-\frac{1}{4\pi}\int_0^{2\pi}\mathrm{d}\varphi \int_{-1}^{+1}\mathrm{d}\ell\,\hat{\bm e}_{\theta}(\ell,\varphi)\sum_{n=1}^\infty p_n\,L_n(\ell)\\
    %&=\frac{\hat{\bm n}}{2}\sum_{n=1}^\infty\frac{2n+1}{n(n+1)}p_n\int_{-1}^{+1}(1-\ell^2)P'_n(\ell)\mathrm{d}\ell\\
    %&=\frac{\hat{\bm n}}{2}\sum_{n=1}^\infty\frac{2n+1}{n(n+1)}\frac{2n(n+1)}{2n+1}p_n\delta_{n1}=
   & =\,\frac{\hat{\bm n}}{D_c}\sum_{m=0}^\infty F_m(\kappa R)\,\frac{m\,\alpha_m}{2m+1}\,\left(\frac{\mu_{m+1}}{2m+3}-\frac{\mu_{m-1}}{2m-1}\right),\\
\end{aligned}
\end{equation}
where $\mu_m$ are defined analogously to $\alpha_m$ via Eq. \eqref{eq:legexp}.
In the last line, we have defined the function $F_m(\kappa R)$ as
\begin{equation}\label{eq:Fm}
   F_m(x)= \frac{(m+1)K_{m+1/2}(x)}{x \,K_{m+3/2}(x)-m\,K_{m+1/2}(x)}.
\end{equation}
For unscreened interactions, $\kappa =0$, the function $F_m$ is unity $F_m(0)=1$ and we retrieve the results in \cite{golestanian2007designing}. For $R\gg 1/\kappa$, i.e. when the radius of the Janus colloid is much larger than the screening length, the self-velocity decays as $1/(\kappa R)$ as $F_m(x)=(m+1)/x+O(1/x^2)$; see Appendix \ref{app:screening} for a detailed discussion.

For a typical half-coated Janus colloid we obtain $\alpha_{2m}=\mu_{2m}=0$ for $m\neq 0$ \cite{golestanian2007designing}, thus
\begin{equation}
    V=-\frac{\alpha_1 \mu_0}{3D_c}F_1(\kappa R),
\end{equation}
where the screening modifies the self-propulsion velocity via the factor $F_1(x)=2(1+x)/(x^2+2x+2)$. By symmetry, the angular velocity $\bm{\omega}=-3\langle \hat{\bm e}_{r}\times\bm{v}_s\rangle|_{r=R}/(2R)=0$ induced by the self-activity is null \cite{anderson1989colloid}. A representation of the flow generated by a single Janus particle is given in Fig.~\ref{fig:schematic} to show a comparison between experimental data [Fig.~\ref{fig:schematic}b] and theoretical predictions [Fig.~\ref{fig:schematic}c].

As outlined in Eqs. \eqref{eq:vstresslet}, \eqref{eq:stokesstresslet}, and \eqref{eq:Sij}, the coefficient $S_{ij}^{\rm in}$ for a half-coated Janus colloid reads
\begin{equation}
\label{singlestresslet}
     S_{ij}^{\rm in} = -\frac{20}{3}\pi\eta R^2 \frac{\alpha_1 \mu_1}{D_c}\phi(\kappa R)\left(n_i n_j-\frac{1}{3}\delta_{ij}\right),%\quad\quad B_j=-\frac{2\pi R^3}{3}\frac{\alpha_1 \mu_0}{D_c}F_1(\kappa R) n_j.
 \end{equation}
 where $\phi(x)$ is a function defined as follows
%\begin{equation}\label{eq:phi}
%\begin{aligned}
%    \phi = &2\sum_{m=1}^\infty F_{m}(\kappa R)\frac{2m+1}{m+1}\left[\int_0^1\mathrm{d}x\,P_{m}(x)\right]\\&\times\left[\int_0^1\mathrm{d}x\,x(1-x^2)P_{m}'(x)\right]\\
%\end{aligned}
%\end{equation}
%\begin{equation}\label{eq:phi}
%\begin{aligned}
%    \phi = &\frac{\sqrt{\pi}}{4}\sum_{l=0}^\infty F_{2l+1}(\kappa R)\left[P_{2l}(0)-P_{2l+2}(0)\right]\\&\times\frac{(2l+1)}{\Gamma\left(\frac{3}{2}-l\right)(l+2)!},\\
%\end{aligned}
%\end{equation}

\begin{equation}\label{eq:phi}
\begin{aligned}
    \phi(x) = &\frac{\sqrt{\pi}}{4}\sum_{l=0}^\infty \frac{(-1)^l (4l+3)}{\Gamma\left(\frac{3}{2}-l\right)(l+2)!}\frac{(2l+1)!!}{(2l+2)!!}\,F_{2l+1}(x),\\
\end{aligned}
\end{equation}
where $\Gamma(u)$ is the Gamma function. %, and ${P_{2l}(0)=(-1)^l (2l-1)!!/(2l)!!}$. 
Screening suppresses the stress terms generated through the factor $\phi(\kappa R)$ which decays as $(\kappa R)^{-1}$ for $\kappa R\gg 1$; see Appendix \ref{app:screening} for a plot.

We note that $S_{ij}^{\rm in}$ contains the nematic order parameter $Q_{ij}= n_i n_j - \delta_{ij}/3$, and is proportional to the first moment of the activity $\alpha_1$, which determines the phoretic self-propulsion $V$. If the colloidal particle has constant and homogeneous surface mobility $\mu(\theta)=\mu_0$ and activity $\alpha(\theta)=\alpha_0$, it does not generate flows at this order of approximation. 

\subsection{Flow due an external chemical gradient}

We now calculate the flow generated by a colloid as a response to an external chemical gradient $\bm{g}=\bm\nabla c_\infty$. In the far field approximation, where we do not consider corrections to the chemical or hydrodynamic field due to the proximity of other particles~\cite{Nasouri2020}, the flow generated by a single active particle in a suspension is the flow produced by it in an imposed chemical gradient. The important difference is that the local chemical gradient is generated collectively and evaluated at the reference particle position $\bm r_0$. In the following discussion, we assume that $\bm r_0$ coincides with the origin of our coordinate system. Consistent with our approximation, we add the flow generated by an external gradient to Eq. \eqref{singlestresslet} to obtain the contribution from a single particle. As in the previous section,
the external gradient $\bm g$ generates the slip velocity $ \bm{v}_s = \mu(\theta)(\mathbb{I}-\hat{\bm e}_{r}\hat{\bm e}_{r})\cdot \bm{g}$; see Eq.~\eqref{eq:vs}. As previously described, the stresslet ${S}_{ij}^{\rm out}$ completely characterizes active flows decaying as $r^{-2}$. For half-coated colloids, Eq.  \eqref{eq:Sij} is explicitly given by,
\begin{equation}
\label{eq:extsource}
    S^{\rm out}_{ij}
    =-2\pi\eta R^2 \mu_1\left(\hat{n}_i g_j+\hat{n}_j g_i-\frac{2}{3}\delta_{ij}\hat{\bm n}\cdot\bm{g}\right).
\end{equation}

%%%%%%%%%%%% COLLECTIVE EFFECTS %%%%%%%%

\section{Collective effects}\label{sec:collective}

To analyze the behavior of the entire suspension, we use the expression for the stress generated by a single active colloid in Eqs. \eqref{singlestresslet} and \eqref{eq:extsource}, accounting separately for self-generated and external chemical gradients. Due to the linearity of the Stokes equation, and the far-field approximation described in the previous section, we can superimpose the individual contributions to obtain the stresses in the suspension. Hence, summing all particles ($\gamma\in\{1,\dots,N_a\}$) and species ($a\in\{1,\dots,N_s\}$), we obtain
\begin{equation}
\begin{aligned}
\label{eq:stokesFull}
&-\eta \nabla^2 v_i + \partial_i P=  \sum_{a=1}^{N_s} \sum_{\gamma =1}^{N_a} (S_{a,ij}^{\rm{out, \gamma}} + S_{a,ij}^{\rm{in, \gamma}}) \partial_j \delta(\mbf r - \mbf r_{a}^\gamma),\\ 
&=  -2 \pi \eta \sum_{a=1}^{N_s} R_a^2 \mu_{a,1} \sum_{\gamma=1}^{N_a}\left(\hat{n}_{a,i}^\gamma \,g_{a,j}^\gamma+\hat{n}_{a,j}^\gamma\, g_{a,i}^\gamma-\frac{2}{3}\delta_{ij}\hat{\bm n}_a^\gamma\cdot\bm{g}_a^\gamma\right.\\  
&\hskip3cm\left.+\frac{10}{3} \frac{\alpha_{a,1}}{D_c}\phi \,Q_{a,ij}^\gamma \right)
\partial_j \delta(\mbf r - \mbf r_{a}^\gamma), 
\end{aligned}
\end{equation}
where $\bm r_{a}^\gamma$ now indicates the position of $\gamma$-th particle of species $a$, namely, where the source is located. Equation \eqref{eq:stokesFull} is the first novel result of our paper: the right-hand side can be viewed as a stress tensor of the fluid expressed in terms of microscopic quantities, i.e. activity and mobility of the half-coated colloids fully characterizing how the suspension of Janus particles modifies the Stokesian fluid. Further simplification to \eqref{eq:stokesFull} comes from the assumption that the number of particles for each species is sufficiently large such that we can substitute the right-hand side with its average with respect to the microscopic noise.
In this regard, we define the following averaged quantities
\begin{equation}
\begin{aligned}
\rho_a&=\Bigg\langle\sum_{\gamma=1}^{N_a}\delta(\bm{r}-\bm{r}_{a}^\gamma)\Bigg\rangle,\;\;\;\\
\bm p_a&= \Bigg\langle\sum_{\gamma=1}^{N_a}\hat{\bm{n}}_{a}^\gamma\,\delta(\bm{r}-\bm{r}_a^\gamma)\Bigg\rangle,\\
\mathbb{Q}_a&= \Bigg\langle\sum_{\gamma=1}^{N_a}\left(\hat{\bm{n}}_{a}^\gamma\hat{\bm{n}}_{a}^\gamma-\frac{\mathbb{I}}{3}\right)\,\delta(\bm{r}-\bm{r}_{a}^\gamma)\Bigg\rangle,
\end{aligned}
\end{equation}
which correspond to the density, polarization, and nematic field for species $a$; here $\langle\cdot\rangle$ denotes the average over the position and orientational noise in Eq. \eqref{eq:langevin}.
The dynamics of these three quantities are reported in Appendix \ref{sec:CG} in details, where we perform a systematic coarse-graining of Eqs. \eqref{eq:langevin} and \eqref{eq:stokesFull}.

Upon averaging the RHS of Eq.~\eqref{eq:stokesFull}, the Stokes equation for the fluid velocity field becomes
\begin{equation}
\begin{aligned}\label{eq:stresscoarse}
&-\eta\nabla^2 v_i+ \partial_i P=2\pi\eta\partial_j\Sigma_{ij} \\
&=-2\pi\eta \sum_{a=1}^{N_s} R_a^2 \mu_{a,1}\partial_j\left(p_{a,i}\partial_j c+p_{a,j}\partial_i c-\frac{2}{3} \delta_{ij}\bm{p}_a\cdot\bm\nabla c \right.\\
&\hskip3cm\left.+\frac{10}{3}\frac{\alpha_{a,1}}{D_c}\phi\,Q_{a,ij}\right).
\end{aligned}
\end{equation}
These equations involve the polarization and the nematic tensor, whose time evolution is determined by a hierarchy of moments that we close at the third order in gradients of the densities. We note that the expression for the active stress tensor in Eq. \eqref{eq:stresscoarse} is formally analogous to that obtained for active liquid-crystalline emulsions in Ref.~\cite{Cates_Tjhung_2018}. For these emulsions, the stress tensor consists of two pieces, one proportional to the nematic tensor $\mathbb Q$ coming purely from the active contribution, and the other, constructed from the polarity and the molecular field, which has an equilibrium origin. We have a formal correspondence of the latter
if one considers $\nabla c$ as the molecular field, but in our case everything originates from out-of-equilibrium interactions.

As reported in Appendix \ref{sec:closure}, the polarity and the nematic tensor can be expressed in terms of the density fields $\rho_a$ as
\beq\label{eq:pen}
 &&  p_{a,i} = \frac{1}{6D_{r,a}}\left(-V_a \partial_i\rho_a+2\Omega_a\rho_a \partial_i c\right),\\ \nonumber
 &&  Q_{a,ij} = \frac{1}{10 D_{r,a}}\left[\Omega_a\left(p_{a,i}\partial_j c+p_{a,j}\partial_i c-\frac{2}{3}\delta_{ij}\bm{p}_a\cdot\bm\nabla c\right)\right.\\ 
 &&\left.-\frac{V_a}{3}\left(\partial_i p_{a,j}+\partial_j p_{a,i}-\frac{2}{3}\delta_{ij}\bm\nabla\cdot\bm{p}_a\right)\right]\label{eq:qen}.
\eeq
Averaging over the RHS of the equation for $c$ in Eq.~\eqref{eq:c} and keeping terms up to second order in gradients, we can finally substitute $  c = \sum_b\bar{\alpha}_b\rho_b +O(\nabla^2)$
with $\bar{\alpha}_a = \alpha_{a,0}/(D_c\kappa^2)$ in Eq.~\eqref{eq:pen} to have the dynamics of the system only expressed in terms of $\rho_a$ and $\bm{v}$. For the density fields, we recover Eq. \eqref{eq:rhoa} with the following coefficients
 \begin{equation}
 \label{eq:parameters_rho}
     \begin{aligned}
         D^{\rm eff}_a &= D_a + \frac{V_a^2}{6 D_{r,a}},  \\
  \chi_{ab} &= \bar{\alpha}_b \left(  \mu_{a,0} - \frac{V_a \Omega_a}{3 D_{r,a}}\right).
     \end{aligned}
 \end{equation}  
Finally, the stress tensor in Eq. \eqref{eq:stresstensor} is obtained by substituting the expression for $c$ and those in Eqs. \eqref{eq:pen} and \eqref{eq:qen} into Eq. \eqref{eq:stresscoarse}.
We can now directly write down the microscopic dependence on the parameters $\lambda_{ab}$ and $\sigma_{ab}$ of Eq. \eqref{eq:stresstensor} previously discussed in Section \ref{sec:Intro}: the coefficients of the first term are
\begin{equation}
 \label{eq:parameters_stress1}
    \begin{aligned}
      \lambda_{ab} &= \frac{2R_a^2 \mu_{a,1} V_a \phi \Omega_a \bar{\alpha}_b}{27 D_c D_{r,a}^2},
    \end{aligned}
\end{equation}
while the second contribution is
 \begin{equation}
  \label{eq:parameters_stress2}
\begin{aligned}
 \sigma_{ab}(\{\rho\}) &=\frac{1}{12}\left(\frac{V_a \epsilon_a}{D_{r,a} }\bar{\alpha}_b+\frac{V_b \epsilon_b}{D_{r,b} }\bar{\alpha}_a\right)   +\frac{1}{2}\left(\lambda_{ab}+\lambda_{ba}\right)\\
 &-\left(\sum_d \rho_d\, \frac{\Omega_d \epsilon_d }{3 D_{r,d}}\right)\bar{\alpha}_a\bar{\alpha}_b,
 \end{aligned}
 \end{equation}
 %where $\{\rho\}=\{\rho_1, \rho_2,\dots, \rho_{N}\}$ and, for half-coated Janus particles, 
 where
\begin{equation}
  \label{eq:parameters_stress3}
    \begin{aligned}
    % \sigma_{ab}^{(0)} &= \frac{V_a \epsilon_a}{6 D_{r,a} }\bar{\alpha}_b   +\lambda_{ab},\\
  \epsilon_a &= R_a^2 \mu_{a,1} \left( 1 + \frac{2\alpha_{a,1}\phi\Omega_a}{3D_c D_{r,a}} \right).\\
 %\omega_a &=\frac{\Omega_a \epsilon_a }{3 D_{r,a}}.  
    \end{aligned}
\end{equation}
With these parameters, we obtain a comprehensive interpretation of Eq. \eqref{eq:rhoa} and Eq. \eqref{eq:stresstensor} in terms of microscopic details. Finally,  one can show \cite{tucci2024nonreciprocal,PhysRevE.89.062316} the following relations between the phoretic coefficients appearing in Eqs. \eqref{eq:langevin} and the Legendre coefficients $\mu_{a,m}$ and $\alpha_{a,m}$ for half-coated colloids
 \begin{equation}
 \begin{aligned}\label{eq:phoreticParameters}
     &\mu_{a}= \mu_{a,0},\quad \Omega_{a} = -\frac{3}{4 R_a}\mu_{a,1},\\
     &\alpha_a=4\pi R_a^2 \alpha_{a,0}, \quad \nu_a=0.
 \end{aligned}
 \end{equation}
 The equations of motion presented in Eqs.~\eqref{eq:rhoa} and \eqref{eq:stresstensor} and the expressions for all the coefficients appearing in these equations enumerated in Eqs. \eqref{eq:parameters_rho}-\eqref{eq:phoreticParameters} provide a complete description of a collection of Janus colloids for an arbitrary number of species in a suspension. We will now discuss a few specific examples of the implication of this system of equations starting with one species. 

\subsection{One species}

Neglecting pressure-like terms, we see from Eqs. \eqref{eq:stresstensor}, \eqref{eq:parameters_stress2}, and \eqref{eq:parameters_stress3}, that the stress tensor has a similar structure to the one derived in phenomenological models of the class of Model H \cite{RevModPhys.49.435}, namely
\begin{equation}
\begin{aligned}
    \Sigma_{ij} &= -\zeta \left(\partial_i\rho  \,\partial_j\rho - \delta_{ij} \frac{1}{3} |\bm{\nabla}\rho|^2 \right),\\
    \end{aligned}
\end{equation}
where the coefficient $\zeta$ can be expressed in terms of microscopic parameters according to
\begin{equation}\label{eq:StressSingle}
\begin{aligned}
    \zeta &= 2(\lambda-\sigma),\\
    &=\frac{R^2 \mu_1 \bar{\alpha}}{3 D_r}  \left(1 + \frac{2 \alpha_1 \phi \Omega}{3D_c D_{r}}\right) (2 \rho \bar{\alpha} \Omega-V).
\end{aligned}
\end{equation}
It is worth remarking that the expression of $\zeta$ reflects its active nature, with no direct link to surface tension contributions that appear in the density equation similar to \cite{tiribocchi2015active}. This is in contrast to what happens for equilibrium models describing the dynamics of a binary mixture coupled to a fluid, as in model H \cite{RevModPhys.49.435,Cates_Tjhung_2018}. Indeed, $\zeta$ can take both positive and negative values due to the competition between collective effects and the flow generated by individual particles. 

This competition is encoded in the terms within the second set of parentheses in Eq.~\eqref{eq:StressSingle}, which determine whether the system behaves as \textit{extensile} ($\zeta > 0$) or \textit{contractile} ($\zeta < 0$) \cite{lauga2016stresslets}.  
In the extensile case ($\zeta > 0$), particles effectively push the fluid outward along their axis, generating a flow that moves fluid forward and backward relative to their orientation. In contrast, in the contractile case ($\zeta < 0$), they pull the fluid inward.  
Assuming all parameters in Eq.~\eqref{eq:StressSingle} are positive, the nature of the effective flow generated by a colloid is governed by the interplay between the alignment interaction $\Omega$ and the self-propulsion velocity $V$.
This calculation can thus be seen as a bottom-up derivation for this class of scalar wet active systems \cite{tiribocchi2015active}.  

\begin{figure}[t]
    \centering
    \includegraphics[width=0.95\columnwidth]{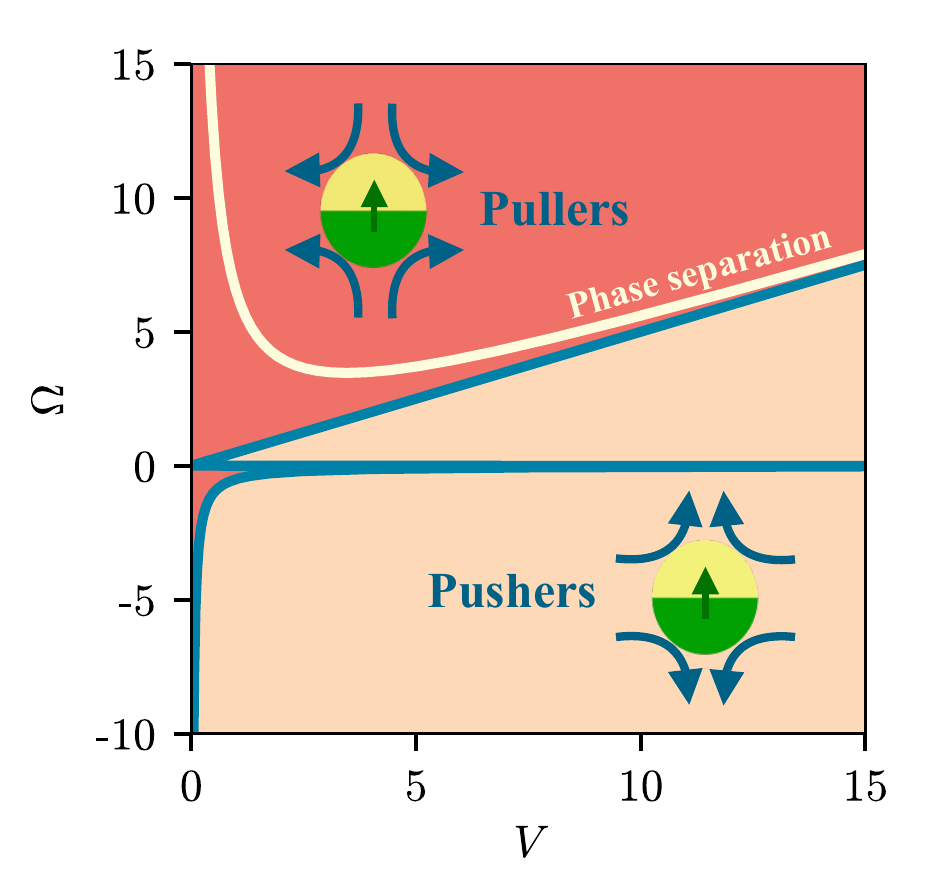}
    \caption{State diagram in the plane of self-propulsion $V$ and alignment $\Omega$ in a suspension of a single species of Janus colloids. The yellow line indicates the threshold for the instability of the homogeneous state $\bar{\rho}$, which is unstable in the region above the line, namely for large positive $\Omega$. The blue line separates the region of positive values of $\zeta>0$ (orange area) where particles are pushers, from where $\zeta<0$ and particles behave as pullers (red area).  The region of instability in the density is associated with a puller's behavior. }
    \label{fig:phase_diagram1sp}
\end{figure}

We report in Fig.~\ref{fig:phase_diagram1sp} the evaluation of Eq.~\eqref{eq:StressSingle} for different values of self-propulsion and alignment fixing the other parameters to unity. We distinguish the region of parameters where particles produce a pusher or puller flow, in connection with a phase-separation instability. For the latter we mean the region of parameters where the solution $\bar{\rho}$ of Eq. \eqref{eq:rhoa} becomes unstable, giving rise to aggregated states \cite{tucci2024nonreciprocal}. To identify it, we assume a linear perturbation of the type $\rho=\bar\rho+\delta\rho$ and $\bm{v}=\delta\bm{v}$, namely a null background fluid velocity.
In linear order of the perturbation, Eq. \eqref{eq:rhoa} is not affected by the fluid and reduces to
\begin{equation}\label{eq:linearrho}
\begin{aligned}
     \partial_t\delta\rho&=\Lambda \nabla^2 \delta \rho \\
    \Lambda  &=D+\frac{V^2}{6\,D_{r}}+\bar{\rho}\bar\alpha\left(\mu-\frac{V\Omega}{3\,D_{r}}\right) \ .
\end{aligned}
\end{equation}
When $\Lambda<0$, the particles are effectively attracted by each other \cite{ouazan2021non}, and it is reasonable to expect aggregated states in this regime. We notice that the region of instability for a homogeneous state is associated with a puller behavior \cite{Jibuti2014Dec, Garcia2013Mar}.

\subsection{Many species}  
For more than one species, the phoretic stress tensor takes on the following structure:
\begin{equation}
\label{eq:stresstensormulti}
    \begin{aligned}
        \Sigma_{ij} &= -\sum_{ab} \partial_i \rho_a  (Z^{\rm{r}}_{ab}+Z^{\rm{nr}}_{ab})\, \partial_j \rho_b, 
        \end{aligned}
\end{equation}
where we identify the symmetric coupling matrix $Z^{\rm r}_{ab}$ between species and the non-reciprocal one $Z^{\rm nr}_{ab}$, given by
\begin{equation}
\label{eq:decomposition}
    \begin{aligned}
        Z^{\rm{r}}_{ab} &= 2 (\lambda^{\rm r}_{ab} - \sigma_{ab}),\quad% = \frac{1}{2}(k_a \bar\alpha_b+k_b \bar\alpha_a), \\
        Z^{\rm{nr}}_{ab} =  \lambda_{ab} - \lambda_{ba}.
    \end{aligned}
\end{equation}
In transitioning from Eq. \eqref{eq:stresstensor} to Eq. \eqref{eq:stresstensormulti} we have discarded pressure-like terms by rewriting $\rho_a\partial_i\partial_j\rho_b=\partial_i(\rho_a\partial_j\rho_b)-\partial_i\rho_a\partial_j\rho_b$. This procedure results in a stress tensor that is no longer explicitly symmetric in the spatial components. However, it does not alter the conservation of total angular momentum but highlights multi-species anti-symmetric contributions. For two species, this simply reduces to
\begin{equation}
    \begin{aligned}
         Z^{\rm{nr}}_{ab}=2\lambda^{\rm nr}\epsilon_{ab}, \quad \lambda^{\rm{nr}} =( \lambda_{12}- \lambda_{21})/2,
    \end{aligned}
\end{equation}
where $\epsilon_{ab}$ denotes the two-dimensional Levi-Civita tensor.
If $\lambda_{12} \neq \lambda_{21}$, as we expect for two different catalytic particles, non-reciprocity emerges and affects the dynamics also through the fluid flow. Therefore, in addition to a direct non-reciprocal interaction between the two species (Eq. \eqref{eq:rhoa}), the system is characterized by long-range non-reciprocal interactions mediated by the fluid. It is illuminating to examine this statement using the derived expression of $\lambda_{ab}$ and assuming $R_1=R_2=R$ and $D_{r,1}=D_{r,2}= D_r$. We obtain
\beq
\lambda^{\rm{nr}} = \frac{ R \phi}{81 \pi D_c D_r^2} (V_1 \Omega_1^2 \bar\alpha_{2} - V_2 \Omega_2^2 \bar\alpha_{1}).
\eeq 
For generic surface activities and mobilities of the two species, this quantity is different from zero.  
Furthermore, by making the additional assumption that the first moments of mobility and activity are equal for the two species, we obtain
\beq
 \lambda^{\rm{nr}} \propto (\mu_{1,0} \alpha_{2,0} -  \mu_{2,0} \alpha_{1,0}),
\eeq 
which is the phoretic non-reciprocity (as introduced in \cite{soto2014,agudo2019active}) that characterizes the non-equilibrium dry dynamics \cite{agudo2019active,saha2020scalar,tucci2024nonreciprocal}.

\section{Conclusions}\label{sec:concl}

A mixture of Janus colloids serves as an ideal playground for the emergence of non-reciprocal dynamics~\cite{tucci2024nonreciprocal}. At the continuum level, that is at the largest time and length scales, it has been shown that a mixture of chemotactic Janus colloids is well described by the non-reciprocal Cahn-Hilliard model~\cite{saha2020scalar}. To probe these predictions in experimental systems it is important to consider the role of fluid flows~\cite{pisegna2025}. In this paper we have derived the equations for the dynamics of conserved densities coupled with a simple fluid satisfying the Stokes equation, illustrating the microscopic origin of the stress terms that are generally written on the basis of symmetry. 

We have highlighted the importance of screening in determining the strength of the stress terms. For a system that contains only one species, we have shown that by tuning the average density and the ratio of the phoretic parameters, the coupling to the stress can acquire either sign---and the fluid flow either contractile or extensile---thus providing a bottom-up approach to Model-H type dynamics~\cite{tiribocchi2015active}. For multi-species suspensions, we showed how phoretic effects generate non-reciprocal couplings in the stress tensor. 

The insights gained from this study are crucial for the future development of wet non-reciprocal phenomenological models and for advancing our understanding of the hydrodynamics underpinning multi-species active field theories. Our bottom-up approach, which bridges microscopic particle dynamics with macroscopic fluid behavior, offers a robust framework for interpreting the collective behavior of Janus colloids. We believe this work not only provides a solid theoretical foundation but also opens new avenues for experimental applications of Janus colloids in a solvent. 

%We have investigated a microscopic model for a suspension of multi-species Janus colloids, in which the self-propelling particles exhibit chemotactic responses to solute gradients while being advected by a viscous fluid. Beginning with the dynamics of a single particle, we examined how its motion influences the surrounding fluid environment. This analysis was then extended to the collective dynamics of the colloids in the far-field approximation. By systematically coarse-graining the colloidal dynamics, we derived the hydrodynamic equations governing the system and formulated an effective stress tensor for the fluid, which satisfies the Stokes equation. Importantly, this allowed us to unveil the dependence of the stress tensor couplings on phoretic microscopic parameters, which can be tuned to generate distinct flow patterns.
%This calculation provides an interesting insight into the large-scale behavior of active suspensions, specifically in relation to the structural properties of the colloids. In the single-species case, our approach offered a micro-to-macro derivation of scalar wet active field theories . 

\section{Acknowledgements}
We thank Navdeep Rana and Yuto Hosaka for discussions. This work has received support from the Max Planck School Matter to Life and the MaxSyn-Bio Consortium, which are jointly funded by the Federal Ministry of Education and Research (BMBF) of Germany, and the Max Planck Society.
 \newline

\bibliographystyle{apsrev4-2}  
\bibliography{bibliography.bib} 
\onecolumngrid

\setcounter{secnumdepth}{3}
\appendix

%\section{Poisson equation with screening}

\section{Coarse-graining}\label{sec:CG}

Here, we discuss the derivation of Eqs. \eqref{eq:rhoa},\eqref{eq:pen} and \eqref{eq:qen} by coarse-graining the $3-$dimensional microscopic dynamics in Eqs. \eqref{eq:langevin} \cite{golestanian2012,tucci2024nonreciprocal,PhysRevE.89.062316,PhysRevResearch.3.013100}. 
We start by considering the evolution of $\mathcal{P}_a(\bm{r},\hat{\bm{n}},t)$, the probability density of a particle of species $a$ to be at position $\bm{r}$ and self-orientation $\hat{\bm{n}}$ at time $t$, defined as
\begin{equation}
\mathcal{P}_a(\bm{r},\hat{\bm{n}},t)=\Bigg\langle\sum_{\gamma=1}^{N_a}\delta(\bm{r}-\bm{r}_{a}^\gamma)\,\delta(\hat{\bm{n}}-\hat{\bm{n}}_{a}^\gamma)\Bigg\rangle,
\end{equation}
for a set of $N_a$ Janus colloids of species $a$, where the average is taken with respect to the noise contribution to the dynamics.
Then, considering that the process in Eq. \eqref{eq:langevin} follows Stratonovich convention of stochastic calculus, it can be easily shown that $\mathcal{P}_a(\bm{r},\hat{\bm{n}},t)$ satisfies the following Fokker-Planck equation
\begin{equation}\label{eq:FP}
\begin{aligned}
\partial_t \mathcal{P}_a(\bm{r},\hat{\bm{n}},t)=&-\bm\nabla\cdot\left\{\left[V_a\hat{\bm{n}}+\bm{v}-\mu_a\bm\nabla c+\nu_a(\hat{\bm{n}}_a\hat{\bm{n}}_a-\mathbb{I}/3)\cdot\bm\nabla c\right]\mathcal{P}_a(\bm{r},\hat{\bm{n}},t)\right\}+D_a\nabla^2\mathcal{P}_a(\bm{r},\hat{\bm{n}},t)\\
%&-\mathcal{R}\cdot\left[\Omega_a\left(\hat{\bm{n}}\times \bm\nabla c\right)\mathcal{P}_a(\bm{r},\hat{\bm{n}},t)\right]+D_{r,a}\mathcal{R}^2\,\mathcal{P}_a(\bm{r},\hat{\bm{n}},t),
&-\bm\nabla_{\hat{\bm{n}}}\cdot\left\{\left[\Omega_a\left(\mathbb{I}-\hat{\bm{n}}_a\hat{\bm{n}}_a\right)\cdot \bm\nabla c+\mathbb{W}\cdot\hat{\bm{n}}_a\right]\mathcal{P}_a(\bm{r},\hat{\bm{n}},t)\right\}+D_{r,a}\mathcal{R}^2\,\mathcal{P}_a(\bm{r},\hat{\bm{n}},t),
\end{aligned}
\end{equation}
where $\mathcal{R}_i\equiv \epsilon_{ijk}n_j\partial_k^{(n)}$ is the orientational gradient operator. It satisfies the following relations $\mathcal{R}_in_j=-\epsilon_{ijk}n_k$ and $\mathcal{R}^2n_i=-2n_i$.
The first line on the right-hand side of Eq. \eqref{eq:FP} describes the contribution to the probability flux due to drift and diffusion of the particle position, whereas the second line refers to the alignment interaction and diffusion of its orientation.
%For our purposes, we assume that the dynamics of the substrate density $c$ is given by
%\begin{equation}
%    \partial_t c = D_c(\nabla^2-\kappa^2)c+\sum_a\alpha_a\rho_a,
%\end{equation}
%where $D_c$ denotes the diffusivity of the chemicals, $\kappa^{-1}$ a degradation/screening length scale, and the last term expresses the fact that Janus colloids produce/consume the chemical with rate $\alpha$.

\paragraph{Density field.---}We restrict our analysis to the first and second moments of the orientation $\hat{\bm{n}}$ by closing the corresponding hierarchy of infinite many equations for the moments generated from Eq. \eqref{eq:FP}.
We start by integrating Eq. \eqref{eq:FP} with respect to $\hat{\bm{n}}$, which leads to the time evolution of density of the particles $\rho_a(\bm{r},t)\equiv\int_{\hat{\bm{n}}}\mathcal{P}_a(\bm{r},\hat{\bm{n}},t)$ of species $a$ 
\begin{equation}\label{eq:rho}
\partial_t\rho_a=-\bm\nabla\cdot\left[V_a\bm{p}_a+\bm{v}-\mu_a\rho_a\bm\nabla c+\nu_a\mathbb{Q}_a\cdot\bm\nabla c\right]+D_a\nabla^2\rho_a ,
\end{equation}
where $\bm{p}_a(\bm{r},t)\equiv\int_{\hat{\bm{n}} }\hat{\bm{n}}\,\mathcal{P}_a(\bm{r},\hat{\bm{n}},t)$ is the polar field and $\mathbb{Q}_a(\bm{r},t)=\int_{\hat{\bm{n}}}\left(\hat{\bm{n}}\hat{\bm{n}}-\mathbb{I}/3\right)\mathcal{P}_a(\bm{r},\hat{\bm{n}},t)$ the nematic tensor (traceless and symmetric).

\paragraph{Polarization field.---}Similarly, one can calculate the dynamics of the polarization field $\bm{p}_a(\bm{r},t)$%\equiv\int_{\hat{\bm{n}} }\hat{\bm{n}}\,\mathcal{P}_a(\bm{r},\hat{\bm{n}},t)$
as
\begin{equation}\label{eq:pippo}
\begin{aligned}
\partial_t p_{a,i}=&-V_a\partial_j Q_{a,ij}-\frac{V_a}{3}\partial_i\rho_a+\left(\mu_a+\frac{\nu_a}{3}\right)\partial_j(p_{a,i}\partial_j c)-\nu_a\partial_j\left(\langle n_i n_j n_k\rangle_{\hat{\bm{n}}_a}\partial_k c\right)+D_a\nabla^2 p_{a,i}\\
&+W_{ij}p_{a,j}-\partial_j(v_j p_{a,i})+\Omega_a\left(\frac{2}{3}\rho_a\partial_i c-Q_{a,ij}\partial_j c\right)-2D_{r,a}p_{a,i},
\end{aligned}
\end{equation}
where $\langle \cdots\rangle_{\hat{\bm{n}}_a}=\int_{\hat{\bm{n}} }\cdots\,\mathcal{P}_a(\bm{r},\hat{\bm{n}},t)$. 
The third-moment tensor $\langle n_i n_j n_k\rangle_{\hat{\bm{n}}_a}$ contains information about lower ones in its trace, hence, it is convenient to rewrite it as the sum of its trace and its traceless component $Q_{a,ijk}^{(3)}$, i.e.,

\begin{equation}\label{eq:Q3}
\langle n_i n_j n_k\rangle_{\hat{\bm{n}}_a}=Q_{a,ijk}^{(3)}+\frac{1}{5}(p_{a,i}\delta_{jk}+p_{a,j}\delta_{ik}+p_{a,k}\delta_{ij}),
\end{equation}
that substituted in Eq. \eqref{eq:pippo} leads to

\begin{equation}\label{eq:piq3}
\begin{aligned}
\partial_t p_{a,i}=&-V_a\partial_j Q_{a,ij}-\frac{V_a}{3}\partial_i\rho_a+\mu_a\partial_j(p_{a,i}\partial_j c)+D_a\nabla^2 p_{a,i}-\nu_a\partial_j(Q_{a,ijk}^{(3)}\partial_k c)-\frac{\nu_a}{5}\left[\bm\nabla(p_{a,j}\bm\nabla c)\right]^{TS}_{ij}\\
&+W_{ij}p_{a,j}-v_j\partial_j p_{a,i}+\Omega_a\left(\frac{2}{3}\rho_a\partial_i c-Q_{a,ij}\partial_j c\right)-2D_{r,a}p_{a,i}.
\end{aligned}
\end{equation}
In the equation above we have introduced the symmetric traceless tensor $[\bm{v} \bm{w}]_{ij}^{TS}$ built from any two vectors $\bm{v}$ and $\bm{w}$ as
%\begin{equation}
$[\bm{v}\bm{w}]^{TS}_{ij}=v_{i}w_{j}+w_{i}v_{j}-2\delta_{ij}v_{k}w_{k}/3.$
%\end{equation}
If one extends formally this definition to differential operators and tensors, we get that the term $\left[\bm\nabla(\bm\nabla c)p_{a,j}\right]^{TS}_{ij}$ in Eq. \eqref{eq:piq3} becomes

\begin{equation}
    \left[\bm\nabla(p_{a,j}\bm\nabla c)\right]^{TS}_{ij}=\partial_i[(\partial_j c)p_{a,j}]+\partial_j[(\partial_i  c)p_{a,j}]-\frac{2}{3}\partial_k[(\partial_k c)p_{a,i}],
\end{equation}
where we recall that we always assume summation over repeated indices.

\paragraph{Nematic field.---}Following the same steps as above, we find that the equation for the nematic field is given by

\begin{equation}\label{eq:Q}
\begin{aligned}
    \partial_t Q_{a,ij}=&\frac{\delta_{ij}}{3}\left[V_a\bm\nabla\cdot\bm{p}_a+\frac{\nu_a}{3}\partial_k(\rho_a\partial_k c)\right]+\mu_a\partial_k(Q_{a,ij}\partial_k c)+D_a\nabla^2 Q_{a,ij}+\frac{\nu_a}{3}\partial_k\left[(\delta_{ij} Q_{a,km}+\delta_{km}Q_{a,ij})\partial_m c\right]\\
    &-\partial_k(u_k Q_{a,ij})+W_{ik}Q_{kj}+W_{jk}Q_{ki}-V_a\partial_k\langle n_i n_j n_k\rangle_{\hat{\bm{n}}_a}-\nu_a\partial_l\left(\langle n_i n_j n_k n_l\rangle_{\hat{\bm{n}}_a}\partial_k c\right)\\
    &+\Omega_a\left(p_{a,i}\partial_j c+p_{a,j}\partial_i c-2\langle n_i n_j n_k \rangle_{\hat{\bm{n}}_a}\partial_k c\right)-6\,D_{r,a}Q_{a,ij},
\end{aligned}
\end{equation}
In order to derive Eq. \eqref{eq:Q} we have calculated the following integrals:
\begin{itemize}
    \item[i)] The first comes from the $\nu$
 interaction and it deals with fourth order moments in $\hat{\bm{n}}$, that is 
 \begin{equation}
    \begin{aligned}
    &-\partial_l\left[(\partial_k c)\Bigg\langle\left(n_i n_j-\frac{\delta_{ij}}{3}\right)\left(n_l n_k-\frac{\delta_{lk}}{3}\right)\Bigg\rangle_{\hat{\bm{n}}_a}\right]\\
    &=-\partial_l\left[(\partial_k c)\langle n_i n_j n_k n_l\rangle_{\hat{\bm{n}}_a}\right]+\frac{\delta_{ij}}{3}\partial_l[(\partial_k c)Q_{a,lk}]+\frac{1}{3}\partial_k[(\partial_k  c)Q_{a,ij}]-\frac{\delta_{ij}}{9}\partial_k[(\partial_k c)\rho_{a}].
    \end{aligned}
    \end{equation}
    
    \item[ii)] The second comes from the alignment interaction proportional to $\Omega_a$
    
    \begin{equation}
        \begin{aligned}
            %-\int_{\hat{\bm{n}}}n_in_j\,\mathcal{R}\cdot[(\hat{\bm{n}}\times\bm\nabla c)\mathcal{P}_a(\bm{r},\hat{\bm{n}},t)]
            &-\partial_k c\int_{\hat{\bm{n}}}\left(n_in_j-\frac{\delta_{ij}}{3}\right)\partial_l^{(n)}\left[(\delta_{lk}-n_ln_k) \mathcal{P}_a\right]\\
            &=\partial_k c\int_{\hat{\bm{n}}}\left(\delta_{il}n_j+\delta_{jl}n_i\right)(\delta_{lk}-n_ln_k) \mathcal{P}_a=p_{a,i}\partial_j c+p_{a,j}\partial_i c-2(\partial_k c)\langle n_i n_j n_k\rangle_{\hat{\bm{n}}_a}.
        \end{aligned}
    \end{equation}
%where we have used the relation $\mathcal{R}_l(n_i n_j)=-n_k(\epsilon_{ljk}n_i+\epsilon_{lik}n_j)$.

\item[iii)] The third one captures the angular diffusion contribution to the dynamics of the nematic tensor
\begin{equation}
        \begin{aligned}
            \int_{\hat{\bm{n}}}\left(n_in_j-\frac{\delta_{ij}}{3}\right)\,\mathcal{R}^2\,\mathcal{P}_a(\bm{r},\hat{\bm{n}},t)=-6\,Q_{a,ij}.
        \end{aligned}
    \end{equation}
where we have used the relation $\mathcal{R}^2n_i n_j = -6(n_i n_j - \delta_{ij}/3)$.
\item[iv)] The last term quantifies the effect of particle rotation due to the fluid:

\begin{equation}
        \begin{aligned}
            &\int_{\hat{\bm{n}}}\left(n_in_j-\frac{\delta_{ij}}{3}\right)\,\partial_k^{(n)}(\delta_{kl}-n_kn_l)W_{lm}n_m\mathcal{P}_a(\bm{r},\hat{\bm{n}},t)=W_{jm}Q_{mi}+W_{im}Q_{mj},
        \end{aligned}
    \end{equation}
where we exploit the anti-symmetry of $W_{ij}$ and the symmetry of $Q_{ij}$.
\end{itemize}

Also in this case we can extract the trace contribution from the fourth moment tensor $\langle n_i n_j n_k n_l\rangle_{\hat{\bm{n}}_a}$ as

\begin{equation}\label{eq:Q4}
\begin{aligned}
    \langle n_i n_j n_k n_l\rangle_{\hat{\bm{n}}_a}=&\,Q^{(4)}_{a,ijkl}+\frac{1}{7}\left(Q_{a,ij}\delta_{kl}+Q_{a,ik}\delta_{jl}+Q_{a,il}\delta_{jk}+Q_{a,jk}\delta_{il}+Q_{a,jl}\delta_{ik}+Q_{a,lk}\delta_{ij}\right)\\
    &+\frac{1}{15}\rho_a(\delta_{ij}\delta_{kl}+\delta_{ik}\delta_{jl}+\delta_{il}\delta_{jk}),
\end{aligned}
\end{equation}
where $Q^{(4)}_{a,ijkl}$ is a traceless tensor.
By substituting Eq. \eqref{eq:Q3} and \eqref{eq:Q4} in \eqref{eq:Q}, we get
\begin{equation}\label{eq:Qiq34}
\begin{aligned}
    \partial_t Q_{a,ij}=&-V_a\partial_k Q_{a,ijk}^{(3)}-\frac{V_a}{5}[\bm\nabla\bm{p}_a]^{TS}_{ij}+\mu_a\partial_k(Q_{a,ij}\partial_k c)+D_a\nabla^2 Q_{a,ij}-u_k\partial_k Q_{ij}+W_{ik}Q_{kj}+W_{jk}Q_{ki}\\
    &-\frac{\nu_a}{7}\left\{[\bm\nabla(\mathbb{Q}_a\cdot\bm\nabla c)]^{TS}_{ij}+\bm\nabla\cdot[\bm\nabla c\,\mathbb{Q}_a]_{ij}^{TS}\right\}-\nu_a\partial_l(Q_{a,ijkl}^{(4)}\,\partial_k c)-\frac{\nu_a}{15}[\bm\nabla(\rho\bm\nabla c)]^{TS}_{ij}-\frac{\nu_a}{21}\partial_k(Q_{ij}\partial_k c)\\
    &+\Omega_a\left\{\frac{3}{5}[\bm{p}_a\bm\nabla c]_{ij}^{TS}-2 Q_{a,ijk}^{(3)}\partial_k c\right\}-6\,D_{r,a}Q_{a,ij}.
\end{aligned}
\end{equation}

\subsection{Closure}\label{sec:closure}

To get a closure for the hierarchy of moments equation we propose a gradient expansion that describes the slow modes of the theory. 
Because of the presence of diffusion in the system, it can be checked that $\partial_t = O(\nabla^2)$, $\bm{p}=O(\bm\nabla)$ and $\mathbb{Q}=O(\nabla^2)$.
Moreover, we make the assumption that $\mu_{m,a}=0$ and $\alpha_{m,a}=0$ for $m\ge 2$, hence $\nu_a=0$, and corresponding to hemispherically coated Janus colloids.
By considering the leading in the gradient expansion  in Eq. \eqref{eq:Qiq34} we get an explicit equation of the nematic tensor in terms of the polarity and density field
\begin{equation}\label{eq:Qexp}
    \mathbb{Q}_a = -\frac{1}{30 D_{r,a}}\left\{V_a[\bm\nabla\bm{p}_a]^{TS}-3\,\Omega_a[\bm{p}_a\bm\nabla c]^{TS}\right\}+O(\bm\nabla^3).
\end{equation}

The equation above can be closed once we express the polar field $\bm{p}_a$ and the chemical density $c$ in terms of $\rho_a$ and its gradients.
Combining equation \eqref{eq:piq3} and \eqref{eq:c} we find

\begin{equation}
\begin{aligned}
    \bm{p}_a &= \frac{1}{6\,D_{r,a}}\left[-V_a\bm\nabla\rho_a+2\Omega_a\rho_a\bm\nabla c\right]+O(\bm\nabla^3),\\[5pt]
    %&= \frac{1}{d(d-1)\,D_{r,a}}\left[-V_a\bm\nabla\rho_a+(d-1)\Omega_a\rho_a\sum_b \bar{\alpha}_b\bm\nabla\rho_b\right]+O(\bm\nabla^3),\\
    c &= \sum_b\bar{\alpha}_b\rho_b+O(\nabla^2),\\
    %&=\sum_b\left\{\bar{\alpha}_b\rho_b+\left[\bar\gamma_b+\frac{V_b\bar{\beta}_b}{d(d-1)\,D_{r,b}}\right]\nabla^2\rho_b-\frac{\Omega_b\bar{\beta}_b}{d\,D_{r,b}}\sum_d\bar{\alpha}_d \bm\nabla\cdot(\rho_b\bm\nabla\rho_d)\right\}+O(\bm\nabla^4),
\end{aligned}
\end{equation}
with $\bar{\alpha}_a = \alpha_a/(D_c\kappa^2)$, and where we have expanded the coarse-grained version of Eq. \eqref{eq:c}. %, $\bar\gamma_a = (\gamma_a+\alpha_a/\kappa^2)/(D_c\kappa^4)$,  $\gamma_a = \alpha_a R_a^2/6$, $\bar{\beta}_a=\beta_a/(D_c\kappa^2)$.
It follows that terms appearing in the nematic tensor in Eq. \eqref{eq:Qexp} become

\begin{equation}
\begin{aligned}
   [\bm\nabla\bm{p}_a]^{TS}&=\frac{1}{6\,D_{r,a}}\left\{-V_a[\bm\nabla\bm\nabla\rho_a]^{TS}+2\Omega_a\sum_b\bar{\alpha}_b[\bm\nabla(\rho_a\bm\nabla\rho_b)]^{TS}\right\}+O(\bm\nabla^4)\\
   &=\frac{1}{6\,D_{r,a}}\left\{-V_a[\bm\nabla\bm\nabla\rho_a]^{TS}+2\Omega_a\sum_b\bar{\alpha}_b\left(\rho_a[\bm\nabla\bm\nabla\rho_b]^{TS}+[(\bm\nabla\rho_a)(\bm\nabla\rho_b)]^{TS}\right)\right\}+O(\bm\nabla^4),\\
   [\bm{p}_a\bm\nabla c]^{TS}
   &=\frac{1}{6\,D_{r,a}}\left\{-V_a\sum_d\bar{\alpha}_d[(\bm\nabla\rho_a)(\bm\nabla\rho_d)]^{TS}+2\Omega_a\sum_{bd}\bar{\alpha}_b\bar{\alpha}_d\,\rho_a[(\bm\nabla\rho_b)(\bm\nabla\rho_d)]^{TS}\right\}+O(\bm\nabla^4).
\end{aligned}
\end{equation}
The equation for the density $\rho_a$ in Eq. \eqref{eq:rhoa} becomes
\begin{equation}
    \partial_t\rho_a = \bm{v}\cdot\bm\nabla\rho_a+\left(D_a+\frac{V_a^2}{6\,D_{r,a}}\right)\nabla^2\rho_a+\left(\mu_a-\frac{V_a\Omega_a}{3\,D_{r,a}}\right)\sum_b\bar{\alpha}_b\bm\nabla\cdot(\rho_a\bm\nabla\rho_b)+O(\bm\nabla^4),
\end{equation}
where we have used the incompressibility condition $\bm\nabla\cdot \bm{v}=0$ and $\bm{v} = O(\bm\nabla)$.

\section{Screening dependence}\label{app:screening}

\begin{figure}[h]
\centering
\includegraphics[width = 0.45\linewidth]{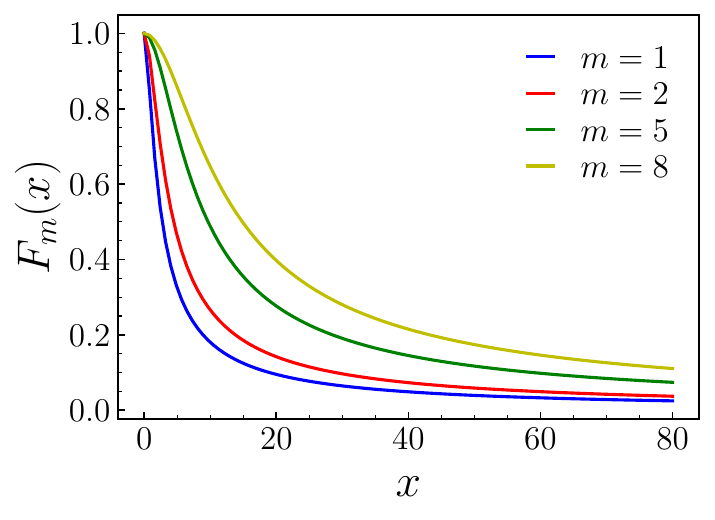}\quad\quad\includegraphics[width = 0.455\linewidth]{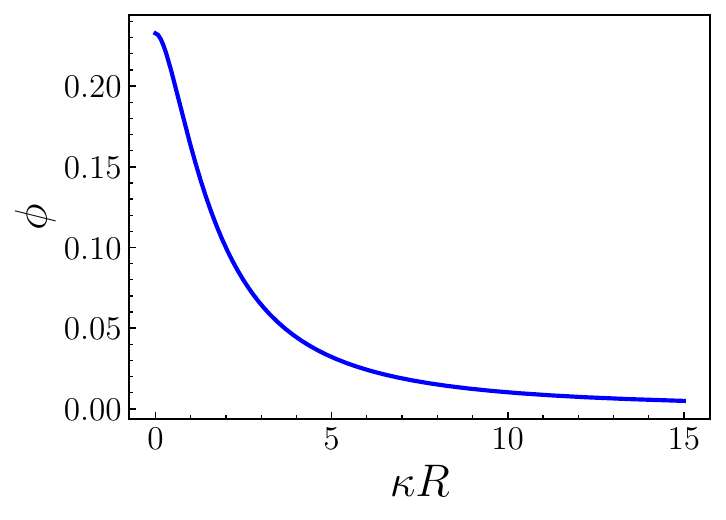}
\caption{On the left, we display $F_m(x)$ as a function of its argument for different value of $m$. On the right, we plot $\phi$ as a function of $\kappa R$. In the absence of screening ($\kappa=0$) it attains the value $\phi \simeq 0.2326 $ \cite{kanso2019phoretic}, while it decays as $(\kappa R)^{-1}$ for $\kappa \gg 1/R$.}\label{fig:fmx}
\end{figure}

As expressed in Eq. \eqref{eq:Ceq}, the presence of the spatial chemical degradation rate $\kappa$ modulates the interactions between different colloids. This is due to the fact that the chemical response of particles at a distance larger than $\kappa^{-1}$ is suppressed exponentially, see e.g. Eq. \eqref{eq:cone}, thus generating a screening effect in the system. Accordingly, also the slip velocity field $\bm v_s$ in Eq. \eqref{eq:vs} and, consequently, the self-propelling velocity in Eq. \eqref{eq:Vsp} depend on the screening via the coefficients $F_m(\kappa R)$, defined in Eq. \eqref{eq:Fm}, where $R$ is the radius of the colloid.
In the low screening regime $\kappa\ll R$, $F_m(\kappa R)=1+O((\kappa R)^2)$, while it decays algebraically as $F_m(\kappa R)=(m+1)/(\kappa R)+ O((\kappa R)^{-2})$ for $\kappa R\gg 1$, as screening effects become stronger. We show the properties of $F_m(x)$ on the left panel of Fig.~\ref{fig:fmx} where we plot it for different values of $m$ and as a function of $\kappa R$. This algebraic behavior can be understood by using the properties of the $K_n$ Bessel function \cite{abramovitz1964handbook}. It can be shown that $F_m(x)$ can written as a ratio of suitable polynomials according to
\begin{equation}
\begin{aligned}
    F_m(x) &= \frac{(m+1)f_m(x)}{\,f_{m+1}(x)-m\,f_m(x) },\quad\quad
    f_m(x) = \frac{1}{2^m}\sum_{k=0}^m\frac{(2m-k)!}{k!\,(m-k)!}\,(2x)^{k}.
\end{aligned}
\end{equation}

Screening also affects the stress generated by the colloids on the surrounded fluid, which is more or less localized depending on the magnitude of $\kappa$. For a single half-coated Janus colloid, the related stress in Eq. \eqref{eq:stokesstresslet} depends on screening via the constant $\phi$ in Eq. \eqref{eq:phi}. It is apparent from the right panel of Fig.~\ref{fig:fmx} that already at $\kappa \simeq 2/R$ screening halves the amplitude of $\phi$, hence of the stress tensor generated by a single colloid.

\end{document}